\documentclass[a4paper]{jpconf}
\usepackage{graphicx}
\begin{document}
\title{Properties of some nonlinear Schrodinger equations motivated through information theory}

\author{Liew Ding Yuan and Rajesh R. Parwani}

\address{Department of Physics, National University of Singapore, Kent Ridge, Singapore.}

\ead{parwani@nus.edu.sg}

\begin{abstract}
We update our understanding of nonlinear Schrodinger equations motivated through information theory. In particular we show that a $q-$deformation of the basic nonlinear equation leads to a perturbative increase in the energy of a system, thus favouring the simplest $q=1$ case. Furthermore the energy minimisation criterion is shown to be equivalent, at leading order, to an uncertainty maximisation argument. The special value $\eta =1/4$ for the interpolation parameter, where leading order energy shifts vanish, implies the preservation of existing supersymmetry in nonlinearised supersymmetric quantum mechanics. Physically, $\eta$ might be encoding relativistic effects.
\end{abstract}

\section{Introduction}

We have previously argued, see for example \cite{gal}, that the maximum uncertainty (entropy) method \cite{jaynes}, used widely in several disciplines \cite{dis}, is also a plausible framework to investigate potential nonlinear generalisations of Schrodinger's equation \cite{par2}.

We consider here for simplicity a single particle in one space dimension. The basic nonlinear equation of Ref.\cite{par2} is 
\begin{equation}
i \hbar {\partial \psi \over \partial t} = - {{\hbar}^2 \over 2m} {\partial^2 \psi \over \partial x^2} + V(x) \psi + 
F(p) \psi \, , \label{nsch1}
\end{equation}
with
\begin{eqnarray} 
F(p) &=& Q_{nl} -Q\, , \\
Q_{nl} &=& { {\cal{E}}  \over \eta^4}  \left[ \ln {p \over (1-\eta) p + \eta p_{+} } + 1 - {(1-\eta) p \over (1-\eta) p + \eta p_{+}} - {\eta p_{-} \over (1-\eta) p_{-} + \eta p} \right] \, , \nonumber \\
Q &=& - {{\hbar}^2 \over 2m}  {1 \over \sqrt{p}} {\partial^2 \sqrt{p} \over \partial x^2} \, , \nonumber 
\end{eqnarray}
$p(x) = \psi^{\star}(x) \psi(x)$ and $p_{\pm}(x)  =  p(x \pm \eta L)$. The nonlinearity length scale $L$ and the energy parameter ${\cal{E}}$ are constrained through the relation 
$ {\cal{E}} L^2 = {\hbar}^2  / 4 m  $.
The dimensionless parameter $0 < \eta \le 1$ plays the role of an interpolating parameter and also regularises potential singularities. 

The nonlinear equation preserves a number of important properties of the linear Schrodinger equation, such as the conservation of probability, existence of the usual plane wave solutions and invariance under a scaling of the wavefunction, $\psi \to \lambda \psi$.

\section{Physical meaning of the $\eta$ parameter}

In Ref.\cite{ng1} nonlinear Dirac equations were constructed using an axiomatic approach \cite{par3}. In taking the non-relativistic limit of those equations \cite{ng2} it was found that the potential singularities in the naive limit were regularised if one kept sub-leading relativistic corrections coming from the lower components of the Dirac spinor. This suggests an interpretation in which the $\eta$ parameter in the nonlinear Schrodinger equation (\ref{nsch1}) is encoding relativistic effects present already in the linear quantum theory, while the nonlinearity is parametrised by the scale $L$.

Alternatively, one may proceed as follows. In Ref.\cite{tab} we computed the perturbative effect of the nonlinearity (\ref{nsch1}) on the energy spectrum of the linear theory with smooth external potentials $V(x)$. For an unperturbed state with quantum number $n$, with $N$ nodes in the wavefunction, the result is 
\begin{equation}
\delta E \approx {\frac{\hbar ^{2}|L| \pi}{6m}}\ \sqrt{\eta(1 - \eta)}\ \
(1-4 \eta)
\sum_{p=1}^{N}C_{np}^{2}\, + O(L/a)^2 \, , \label{EPX}
\end{equation}
where $a$ is the characteristic length scale of the linear theory. The coefficients $C_{np}$ depend on the slope of the unperturbed wavefunctions near the nodes, and hence on the form of the potential. Notice however that the dependence of the energy shift on $\eta$ is universal, that is, independent of $V(x)$. 
(For states without nodes the energy shift is of lower order, $O(L/a)^2$).

It was suggested in \cite{gal} that $\eta$ be fixed by minimising the energy of the system.
The result (\ref{EPX}) indicates that this happens when $\eta \approx 0.80$. Now, to leading order, the energy shift (\ref{EPX}) is given by $<F(p)>$, evaluated using the unperturbed states. Assuming integration over an infinite domain, this reduces after a simple shift of variables \cite{par2} to $I_{NL} -I_F$, where $I_F$ is the Fisher information measure that leads to the linear Schrodinger equation \cite{par1, reg} and $I_{NL} =-I_{KL}$ is the information measure that leads to the nonlinear generalisation (\ref{nsch1}); $I_{KL}$ is the Kullback-Liebler measure.

Thus minimising the energy shift is equivalent to minimising the information content of the nonlinear theory relative to that of the linear theory, and this is achieved for excited states at the universal value $\eta \sim 0.8$.

An interesting aspect of the result (\ref{EPX}) is that it vanishes for values $\eta =0,1/4,1$. The $\eta=0$ case is obvious becuse then the nonlinearity is switched off. The $\eta=1$ case is amusing because there the nonlinearity exists but the intermediate perturbative calculations are ill-defined if one starts with $\eta=1$; thus $\eta \to 1$ may only to taken after the calculations. We will see in the next section that the vanishing of the energy shift at $\eta=1$ is likely to be an accident, valid only at leading order.

So what about the $\eta=1/4$ case? This is quite puzzling and we still do not understand if it indicates some hidden symmetry of the theory at that value. However we see that at that special value the nonlinearity can help preserve, to leading order, an existing supersymmetry of the linear theory: Consider one dimensional supersymmetric quantum mechanics \cite{susy} with two partner Hamiltonians $H_1, H_2$. Except for the ground state of $H_1$ the two Hamiltonians are isospectral but with differing wavefunctions. For example, the first excited state of $H_1$ will have the same energy as the ground state of $H_2$; the corresponding wavefunction of $H_1$ will have one node while that for $H_2$ will be nodeless. Thus if we nonlinearise both Hamiltonians as in (\ref{nsch1}), we see that at $\eta=1/4$, all energy shifts, whether for excited or ground states will be of order $(L/a)^2$ and so to leading order the pair of Hamiltonians will have their original supersymmetry preserved.

\section{Robustness of results}

The nonlinearity in Ref.\cite{par2} was obtained using, arguably, the simplest generalisation of the Fisher information measure which preserved various desirable properties of the former. How sensitive are the above results to the particular choice of $I_{KL}$? We can study this issue by considering  deformations of the 
Kullback-Liebler measure. 

Consider therefore the $q-$deformed measure ($q >0$),
 \begin{equation}
M(\eta,q) \equiv {-1 \over q \eta^4 } \int dx \ dt \ p(x) \left[ \ln_{q} {p(x) \over (1-\eta) p(x) + \eta p(x+\eta L)} \right] \, \label{M2},
\end{equation}
with 
\begin{equation}
\ln_{q} y \equiv {y^{q-1} -1 \over q-1} \, .
\end{equation}
This measure reduces, when $q=1$, to the measure $-I_{KL}$ used in the derivation of (\ref{nsch1}) and retains for general $q$ the various desirable properties of $I_{KL}$. Using the more general (\ref{M2}) leads to another nonlinear Schrodinger equation which can be used to calculate leading order energy shifts as in Ref.\cite{tab}. For unperturbed states with nodes in a symmetric potential the correction may be written  
\begin{equation}
\delta E(q) \approx {\frac{\hbar ^{2}|L| }{4m}} J(\eta,q) \sum_{p=1}^{N}C_{np}^{2}\, + O(L/a)^2 \label{EPq}
\end{equation}
where
\begin{eqnarray}
\lefteqn{ J(\eta , q) \equiv } \nonumber \\
& & \int_{-\infty }^{\infty} {dy y^{2} \over \eta^4 q} \mbox{[} 
 \ln_q {\frac{y^{2}}{(1-\eta )y^{2}+\eta (y+\eta )^{2}}}+  \left({\frac{y^{2}}{(1-\eta )y^{2}+\eta (y+\eta )^{2}}} \right)^{q-1}  \nonumber \\
&& \;\;\;\;\;\;\;\;\;\;\;\;\;\;\;\;\;\;  - (1-\eta )\left({\frac{y^{2}}{(1-\eta )y^{2}+\eta (y+\eta )^{2}}}\right)^q - \eta \left({\frac{ (y-\eta )^{2}
}{(1-\eta )(y-\eta )^{2}+\eta y^{2}}}\right)^q \mbox{]} \,.  \nonumber 
\end{eqnarray}

For integer values of $q$ one may evaluate the function $J(\eta,q)$ explicitly but this becomes time consuming for large $q$ values. Thus we evaluated it numericaly for a range of real  $0<q<1$. The results are as follows.

For fixed $q>1$ and $0<\eta<1$: $J$ has one local maximum at $\eta_{max}$ where it is positive, one global minimum at $\eta_{min}>\eta_{max}$ where it is negative, and it becomes zero at three points $0=\eta_0 <\eta_1 <\eta_2 <1$.  $\eta_{max}$ is between zero and $\eta_1$ while $\eta_{min}$ is between $\eta_1$ and $\eta_2$. $J$ diverges to positive infinity as $\eta \to 1$. The values of $\eta_1, \eta_2$ decrease as $q$ increases. The minimum value of $J$ also increases (becomes less negative) as $q$ increases. In the limit $q \to 1$, $\eta_2 \to 1$ and thus $J$ vanishes at $\eta=1$ as we saw in the last section.

For fixed $0<q<1$ and $0<\eta<1$: $J$ has one global maximum at $\eta_{max}$ where it is positive, one global minimum at $\eta_{min}>\eta_{max}$ where it is negative, and it becomes zero at two points $0=\eta_0 <\eta_1 <1$.  $\eta_{max}$ is between zero and $\eta_1$ while $\eta_{min}$ is between $\eta_1$ and $1$. $J$ is negative at $\eta =1$. The value of $\eta_1$ decreases as $q$ increases. The minimum value of $J$  decreases (becomes more negative) as $q$ increases. 

The results of the last two paragraphs show that $J(\eta,q)$ reaches its minimum (which is negative) for $q=1$ and for $\eta=0.8$; see Fig.(1). Thus using the motivation of the last section, minimising the energy of the system, or equivalently minimising our information about the system, leads to the $q=1$ nonlinear equation (\ref{nsch1}) as the preferred choice, at least when analysing static properties of states with definite parity. 

\begin{figure}[h]
\begin{center}
\includegraphics[width=5in]{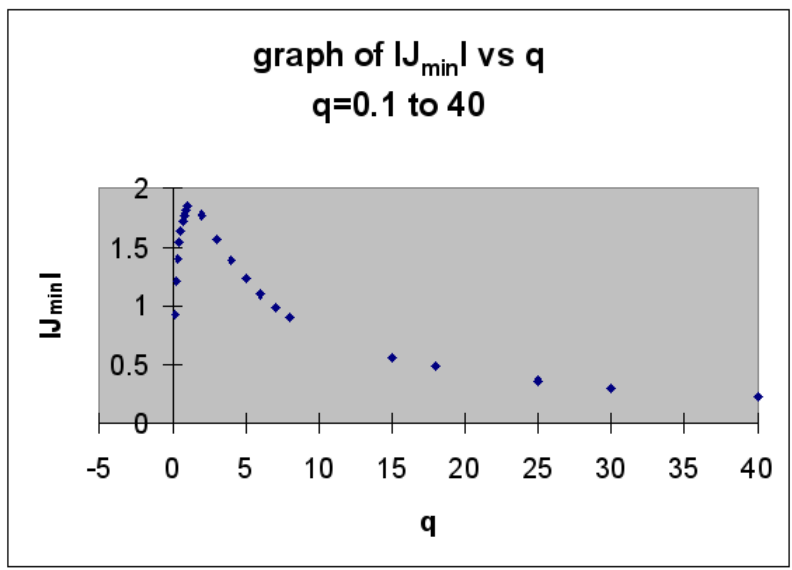}
\end{center}
\caption{The magnitude of the minimum value of $J(\eta,q)$ for each $q$ is plotted}
\end{figure}

So far we have focussed on states with nodes. For states without nodes the energy shifts are $O(L^2)$. 
For the simple harmonic oscillator we have explicitly
\begin{equation}
\delta {E(q)} =  \eta^2 ( \eta^2(q+1)(q+2)- 4\eta(q+1) +2 )
\frac{ \hbar^2 L^2 }{8 m a^4 } + O(L^4) \, .  \label{sho-o}
\end{equation}
This has a  global minimum at $q \sim 1.1$ and $\eta \sim 0.76$, very close to the values for the states with nodes.

\section{Other results and outlook}
Some non-perturbative aspects of the nonlinear Schrodinger equation (\ref{nsch1}) were studied in \cite{RP2}. In \cite{nguyen} we used Eq.(\ref{nsch1}) to study the issue of singularity avoidance in quantum cosmology and further work in this direction is in progress. We have also used information theoretic ideas to construct integrable nonlinear Schrodinger equations in Ref.\cite{PP}. An application of the nonlinear Dirac equations of Ref.\cite{ng1} is in Ref.\cite{ng3}. 

We welcome correspondence from interested readers.

\section*{Acknowledement}
R.P thanks the organisers of the DICE workshop for their hospitality and the opportunity to present our results in a wonderful environment.\\

\end{document}